\documentclass[onecolumn,showpacs,showkeys,preprint,prb]{revtex4-1}

\usepackage{amssymb, amsmath}
\usepackage[english]{babel}
\usepackage{bm}
\usepackage{graphicx}
\usepackage[utf8]{inputenc}
\usepackage{color}

\begin{document}
\author{Marten Richter}
\email[]{marten.richter@tu-berlin.de}

\affiliation{Institut für Theoretische Physik, Nichtlineare Optik und
Quantenelektronik, Technische Universität Berlin, Hardenbergstr. 36, EW 7-1, 10623
Berlin, Germany}
\author{Michael Gegg}
\affiliation{Institut für Theoretische Physik, Nichtlineare Optik und
Quantenelektronik, Technische Universität Berlin, Hardenbergstr. 36, EW 7-1, 10623
Berlin, Germany}
\author{T. Sverre Theuerholz}
\author{Andreas Knorr}
\affiliation{Institut für Theoretische Physik, Nichtlineare Optik und
Quantenelektronik, Technische Universität Berlin, Hardenbergstr. 36, EW 7-1, 10623
Berlin, Germany}

\title{Numerically exact solution of the many emitter -- cavity laser problem: application to the fully quantized spaser emission}

\begin{abstract}
A numerically exact solution to the many emitter -- cavity problem as an open many body system is presented. The solution gives access to the full, nonperturbative density matrix and thus the full quantum statistics and quantum correlations. The numerical effort scales with the third power in the number of emitters. Notably the solution requires none of the common approximations like good/bad cavity limit. As a first application the recently discussed concept of coherent surface plasmon amplification -- spaser -- is addressed: A spaser consists of a plasmonic nanostructure that is driven by a set of quantum emitters. In the context of laser theory it is a laser in the (very) bad cavity limit with an extremely high light matter interaction strength. The method allows us to answer the question of spasing with a fully quantized theory.
 
\end{abstract}


\date{\today}
\maketitle

\section{Introduction}
\begin{figure}
 \includegraphics[clip,width=6cm]{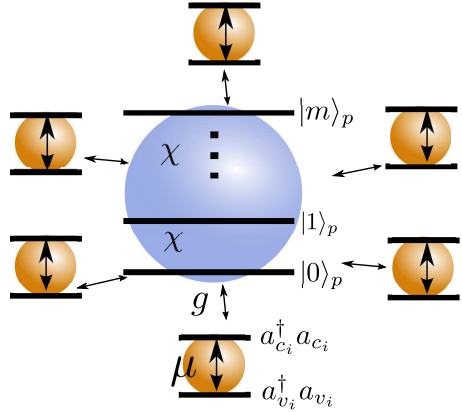}
 \caption{Scheme of the metal nanoparticle/quantum emitter system. The
 system consists of a metal nanoparticle represented by the number states
 $|m\rangle_p$ of a single plasmon mode and a large number of quantum
 emitters with state $v_i$ and $c_i$ for quantum emitter $i$ surrounding
 the metal nanoparticle.}
 \label{levelscheme}
 \end{figure}

For decades open many body quantum systems consisting of a set of many ($N$) externally driven two level quantum emitters (QEs), e.g. dye molecules or quantum dots, coupled to a lossy cavity/optical mode have been subject to extensive research \cite{Carmichael:PhysRevA:86,Drummond:PhysRevA:81,Fink:PhysRevLett:09,Kubo:PhysRevLett:10,Parfenyev:OptExp:14,Protsenko:PhysUsp:12}. These systems provide access to a manifold of interesting physics and real life applications, such as lasers, parametric amplifiers, atomic coherent states. Such model systems have been discussed in the context of quantum computing\cite{Fink:PhysRevLett:09,Kubo:PhysRevLett:10} and quantum plasmonics\cite{Parfenyev:OptExp:14,Protsenko:PhysUsp:12}.
In quantum information processing the coherent exchange of quantum information between QEs and cavity mode requires strong coupling, which can be reached by increasing the number of QEs \cite{Kubo:PhysRevLett:10}. This is desirable, since increasing the emitter numbers allows for a greater parameter range for light matter interaction strength and cavity lifetime at device operating conditions.
In the field of quantum plasmonics, the model system (see Fig \ref{levelscheme}) was utilized to address the feasibility of spasing -- i.e. surface plasmon amplification by stimulated emission of radiation  \citep{Parfenyev:OptExp:14,Protsenko:PhysUsp:12,Bergman:PhysRevLett:03}.

The closed system version of the cavity -- $N$ emitter model is exactly solvable and is known as Tavis-Cummings model \cite{Tavis:PhysRev:170}. The open system counterpart is usually described by a Born-Markov quantum master equation or rather Lindblad equation to include external pumping and losses. There are several approximation schemes for solving the system -- e.g. based on the coherent state positive P representation\cite{Carmichael:PhysRevA:86,Drummond:PhysRevA:81,Parfenyev:OptExp:14} or expansion in an infinite hierarchy of operator expectation values\cite{Protsenko:PhysUsp:12,Gies:PhysRevA:07}.
In this paper we introduce a nonperturbative expansion scheme for the Lindblad equation of many emitters coupled to one optical mode. The method is based on a number state representation. The exponential number of QE degrees of freedom is reduced by assuming identical emitters with identical couplings and dephasings without any use of further approximations. The complexity of the solution then scales with the third power of the emitter number, so that large scale simulations with high emitter numbers are feasible while keeping the full information of the density matrix.

As a first application, the method is applied to the recent topic of coherent plasmon amplification -- spaser. The spaser was introduced by Bergman and Stockman\cite{Bergman:PhysRevLett:03}. The spaser was suggested to provide a coherent source in the emerging field of nanoplasmonics
\cite{Stockman:NewJPhys:08,Ringler:PhysRevLett:08,David:JChemPhys:10,Novotny:NatPhoton:11,Aeschlimann:Science:11,Vasa:ACSNano:10,Lange:Langmuir:12,Kewes:ApplPhysLett:13,Zhang:PhysRevB:14,Richter:PhysRevB:12}. 
A spaser is the surface plasmon analogon of a laser: the cavity is replaced by a metal nanoparticle (MNP) providing bosonic surface plasmon modes, while gain and pump are
completely analogous to classical lasers with active gain medium (atoms, quantum dots). However, claims concerning the experimental realization by Noginov \emph{et al.} (Ref. \onlinecite{Noginov:Nature:09}) have been discussed controversially and questioned, mostly on the basis of a semiclassical theory
\cite{Khurgin:NatPhoton:14,Khurgin:OptExp:12,Zhong:PhysRevB:13,Parfenyev:OptExp:14,Protsenko:PhysUsp:12}. 

Using the fully quantized theory, we confirm that (i) for realistic parameters the spaser behaves like a thresholdless laser, for which the input-output curve cannot be used as indication for spasing and (ii) in a realistic scenario, too high pump rates are required to reach the spaser limit, which is in agreement with the literature using a semiclassical approach \cite{Khurgin:NatPhoton:14,Khurgin:OptExp:12,Zhong:PhysRevB:13}.
Related studies of spasers (semiclassical or single plasmon limit respectively) of a single QE coupled to a plasmon mode were done in Ref. \onlinecite{adrianov1,*adrianov2,*adrianov3}. 
In particular, we present an analysis of the statistics of the created plasmon and exciton distribution to decide for which parameters the system is spasing.
Also, from the calculated full probability distribution, one can distinguish different limits like thermal or coherent plasmon distributions, which determine specific $g^{(2)}$ functions, i.e. plasmon--plasmon correlations.
(The $g^{(2)}$ of the plasmons might be measured from the light emitted from the plasmons of the metal nanoparticles using a Hanburry-Brown-Twiss experiment \cite{Brown:Nature:56,Heeres:NanoLett:2010}.)
Since only the full plasmon statistics determines whether the device is spasing, the calculation of the full probability distribution of the plasmon numbers $\{ p_n \}$ (defined later in detail) as in our method is a major advance.
Previously, the quantum statistics of plasmons of a  metal nanoparticle coupled to one (two) quantum dots were analyzed under resonant excitation \cite{Ridolfo:PhysRevLett:10,Theuerholz:PhysRevB:13}. 
However, to discuss spaser-action a high number of quantum emitters under incoherent off-resonant excitation should be investigated, e.g. something at least in the order of ten or hundred emitters \cite{Protsenko:PhysUsp:12}. Typically the number of emitters required for spasing will depend on the coupling between emitters and plasmons and the various decay channels in the system. In order to get the full information about the system a sophisticated theoretical approach aiming at the full density matrix is necessary. 

\section{Model}

We start with the Tavis-Cummings Hamiltonian\cite{Tavis:PhysRev:170}, i.e. the QEs have identical properties and the same coupling $g$ to the cavity mode.
The free Hamiltonian of emitters and mode takes the form
\begin{eqnarray}
 H_0=\hbar \omega_{sp} b^\dagger b +  \hbar \varepsilon_{c}  \sum_i a^\dagger_{c_i} a_{c_i} +\hbar \varepsilon_{v}  \sum_i a^\dagger_{v_i} a_{v_i}, \label{ham_start}
\end{eqnarray}
with the Boson plasmon creation and annihilation operator $b^\dagger$, $b$, the Fermi creation and annihilation operators
$a^\dagger$, $a$ for the electrons with valence and conduction band levels $v_i$ and $c_i$, the plasmon frequency $\omega_{sp}$ and electron frequencies $\varepsilon_{c}$ and $\varepsilon_{v}$.
We use semiconductor notation\cite{Kira::12} throughout this paper but the Pauli spin matrix notation is easily recovered by setting $\sigma^z_i = a^\dagger_{c_i} a_{c_i} -a^\dagger_{v_i} a_{v_i}$, $\sigma^{+}_i=a^\dagger_{c_i} a_{v_i}$ and $\sigma^{-}_i=a^\dagger_{v_i} a_{c_i}$. The QE -- cavity mode Tavis-Cummings interaction Hamiltonian assumes linear coupling $g$ and rotating wave approximation:
\begin{eqnarray}
 H_{I}=\hbar \sum_ig(  a^\dagger_{v_i} a_{c_i} b^\dagger +  a^\dagger_{c_i} a_{v_i} b).\label{ham_end}
\end{eqnarray}

In the electronic ground state of the emitters $|g\rangle_e$ all electrons are in the confined valence band state, it is constructed from the electron vacuum band state $|vac\rangle_e$ through $|g\rangle_e=\prod_i a^\dagger_{v_i}  |vac\rangle_e$.
The electron states are expanded using exciton states, starting with the single exciton states $|i_1\rangle_e=a^\dagger_{c_{i_1}} a_{v_{i_1}} |g\rangle_e$ with quantum emitter $i_1$ excited, the two exciton states $|i_1, i_2\rangle_e=a^\dagger_{c_{i_2}} a_{v_{i_2}} |i_1\rangle_e$ with quantum emitters $i_1$ and $i_2$ excited.
We can define a general multi exciton state $|\{i_k\}\rangle_e=\prod_{h \in  \{i_k\}} a^\dagger_{c_{h}} a_{v_{h}} |g\rangle_e$ with all quantum emitters $h$ in set $\{i_k\}$ excited.
The cavity mode is described by the plasmon number states $|m\rangle_{p}$.\\
The complete emitter-cavity mode dynamics is described by the density matrix $\rho$.
The Liouville-von Neumann equations together with dissipative corrections $\mathcal{L}$ (line width and relaxation processes in emitters and metal nanoparticle) describe the full system dynamics:
\begin{eqnarray}
 \partial_t \rho = -\frac\imath\hbar [H,\rho]_-+\mathcal{L} \rho . \label{liouville_vonneumann}
\end{eqnarray}
The Lindblad super operator \cite{Breuer::02} has the form $\mathcal{L}\rho=\sum_k \frac{\gamma_i}{2}(2 A_k\rho A^\dagger_k-A^\dagger_k A_k \rho-\rho A^\dagger_k A_k)$.
Typical processes modelled by the super-operator are spontaneous emission of the QEs, phase destroying processes, cavity loss and pumping. Radiative decay of the excitons is described by $A_{E,i}=a^\dagger_{v_i} a_{c_i}$, $\gamma_{E,i}=\gamma_x$; coupling of the plasmon mode to an external Bosonic mode continuum by $A_{sp,1}=b$, $A_{sp,2}=b^\dagger$, $\gamma_{sp,1}=\gamma_{sp}(m+1)$, $\gamma_{sp,2}=\gamma_{sp}m$ with $m=1/(\mathrm{exp}(\hbar\omega_{sp}/(k_B T))-1)$; pure dephasing of the quantum emitter polarizations by $A_{E,i}^{pure}= (a^\dagger_{c_i} a_{c_i}-a^\dagger_{v_i} a_{v_i})$, $\gamma_{E,i}^{pure}=\gamma_{pd}$, and incoherent pumping of the quantum emitters by $A_{E,i}^{pump}= a^\dagger_{c_i} a_{v_i}$, $\gamma_{E,i}^{pump}=P$. The incoherent pump term is known to quench boson output in laser/spaser devices caused by polarization damping, but is nonetheless appealing due to its simplicity \cite{Gartner:PhysRevA:84}.
We will use the spaser example to introduce our method as it makes the presentation more comprehensible. Note that whether these equations describe a laser or a spaser just depends on the parameter domain and on the interpretation whether $b$, $b^{\dagger}$ denote plasmon or photon operators.\\
The matrix elements of the full system density matrix have  the general form 
\begin{equation}
\langle m |_p \langle \{i_k\} |_e \rho | \{i_h\}\rangle_e | m' \rangle_p.
\end{equation}
Since all emitters and couplings are assumed to be identical, it is only important for the density matrix element $\langle m_L |_p \langle \{i_k\} | \rho | \{i_h\}\rangle | m_R \rangle_p$, how many  emitters are excited only in state  $| \{i_h\}\rangle $ and $| \{i_k\}\rangle$ and how many emitters are excited in both states!   
Therefore many matrix elements of the density matrix are identical, this property  reduces the numerical effort for a high numbers of emitters to a feasible level.
We define 
\begin{equation}
\langle m_L |_p \langle \{i_k\} | \rho | \{i_h\}\rangle | m_R \rangle_p=:\rho_{[n_{LR}, n_L, n_R, m_L, m_R ]}
\end{equation} 
with $n_{LR}$ the number of excited emitters  both in the left $\{i_k\}$ and right side $\{i_h\}$   ( $ \{i_k\} \cap  \{i_h\} $) of the density matrix element, $n_L$ the number of excited emitters  only in  the left side $\{i_k\}$  ($ \{i_k\}\setminus  \{i_h\}$) and  $n_R$ the number of elements  only in the right side  ($ \{i_h\}\setminus  \{i_k\}$) of the density matrix element indices.\\
In a first step, the equations of motion are calculated using the von-Neumann equation including dissipators Eq. (\ref{liouville_vonneumann}). In a second step the matrix elements $\langle m_L |_p \langle \{i_k\} | \rho | \{i_h\}\rangle | m_R \rangle_p$ are replaced with $\rho_{[n_{LR}, n_L, n_R, m_L,m_R ]}$ yielding a closed equation of motion system of the form:
\begin{eqnarray}
 &&\partial_t \rho_{[n_{LR}, n_L, n_R, m_L, m_R ]}\nonumber\\
 &&\quad= \imath ( \omega_{sp} (m_R-m_L)+(\varepsilon_c-\varepsilon_v) (n_R-n_L))\rho_{[\dots]}\nonumber\\ 
 &&\qquad+  %\left. \partial_t \rho_{[\dots]}\right|_{L} +
 \left. \partial_t \rho_{[\dots ]}\right|_{xp} +  \left. \partial_t \rho_{[\dots ]}\right|_{diss}.  \label{eq_mo_start}
\end{eqnarray}
In order to simplify the notation, we denote only the indices, which are changed compared to the density matrix written on the lhs of Eq. (\ref{eq_mo_start}).
The emitter-plasmon coupling  causes the formation of (multi-) exciton-plasmon polariton states and resonances.
The contribution of the quantum emitter-plasmon coupling is given by:
\begin{eqnarray}
  &&\left.\partial_t \rho_{[n_{LR}, n_L, n_R, m_L, m_R ]}\right|_{xp}\nonumber\\
  &&\quad = \imath g \{\sqrt{m_R+1} (n_{LR} \rho_{[n_{LR}-1,n_{L}+1,\dots,m_R+1]} \nonumber\\
  &&\qquad \qquad +n_{R} \rho_{[\dots,n_{R}-1,\dots, m_R +1]}) \nonumber\\
  &&\qquad \quad+\sqrt{m_R} (n_L  \rho_{[n_{LR}+1,n_{L}-1,\dots,m_R-1]} \nonumber \\
  &&\qquad \qquad+(N-n_{LR}-n_{L}-n_R) \rho_{[\dots,n_{R}+1,\dots,m_R-1]})  \nonumber\\
  &&\qquad \quad-\sqrt{m_L+1} ( n_{LR} \rho_{[n_{LR}-1,\dots,n_{R}+1,m_L+1,\dots]} \nonumber\\
  &&\qquad \qquad +n_L \rho_{[\dots,n_{L}-1,\dots,m_L+1,\dots]})\nonumber\\
  &&\qquad \quad-\sqrt{m_L} ( n_{R} \rho_{[n_{LR}+1,\dots,n_{R}-1,m_L-1,\dots]}\nonumber\\
    &&\qquad \qquad+(N-n_{LR}-n_{L}-n_R) \rho_{[\dots,n_{L}+1,\dots,m_L-1,\dots]})\}.\label{eq_mo_2}
\end{eqnarray}
The coupled density matrix hierarchy, Eq. (\ref{eq_mo_2}), forms plasmon-polariton states on different excitation levels similar to a Jaynes-Cummings ladder and is known as Tavis-Cummings model.
The large number of different terms arises from rewriting the density matrix elements using the notation $\rho[\dots]$. The underlying processes are of minor complexity, since the action of each interaction is only to either increase or decrease the number of excitations or plasmons on the left or right side (row and column) of the density matrix.
Besides the contributions from the system Hamilton operator Eqs. (\ref{ham_start}-\ref{ham_end}), also the dissipative contributions have to be written in the new formalism:
\begin{eqnarray}
 &&\left.\partial_t \rho_{[n_{LR}, n_L, n_R, m_L, m_R ]}\right|_{diss}\nonumber\\
 &&\quad =\gamma_x\{(N-n_{LR}-n_L-n_R) \rho_{[n_{LR}+1, \dots]} \nonumber\\
 &&\qquad\quad -(2n_{LR}+n_L+n_R)/2 \rho_{[\dots]}\}-\gamma_{pd}(n_L+n_R) \rho_{[\dots]}\nonumber\\
 &&\quad\quad  + P \{ n_{LR} \rho_{[n_{LR}-1, \dots]} - (N - n_{LR} - (n_L+n_R)/2) \rho_{[\dots]})\}\nonumber\\
 && \quad\quad +\gamma_{sp,1}\{\sqrt{(m_L+1)(m_R+1)}   \rho_{[\dots, m_L+1, m_R+1 ]} \nonumber\\
 &&\qquad\quad- (m_L+m_R)/2\rho_{[\dots]}\}\nonumber\\
 && \quad\quad +\gamma_{sp,2}\{\sqrt{m_Lm_R}   \rho_{[\dots, m_L-1, m_R-1 ]}  \nonumber\\
 && \qquad\quad- (m_L+m_R+2)/2\rho_{[\dots]}\}. \label{eq_mo_end}
%%   %TC:incbib
\end{eqnarray}
This gives a complete set of equations to describe the dynamics of the coupled quantum emitters, plasmon system.\\ 
Nondiagonal elements of $\rho_{[\dots ]}$ (i.e. $n_{L}, n_{R}, |m_{L}-m_{R}|>0$) are only included up to numerical convergence.
 \begin{figure}
\includegraphics[clip,width=8.5cm]{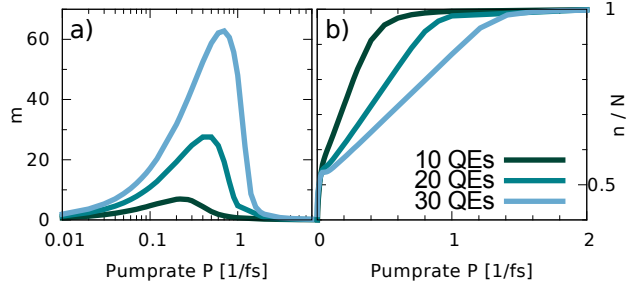}
\caption{(a) Plasmon number expectation value over pump rate and (b)
number of exciton in the quantum emitter system relativ to the number of
emitters. }
\label{inputoutput}
\end{figure}

\begin{figure}
\includegraphics[clip,width=8.cm]{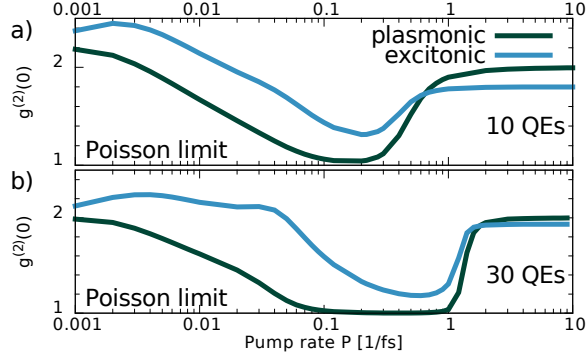}
\caption{The $g^{(2)}$ functions as a function of the pumprate for a)
$N=10$ and b) $N=30$.}
\label{g2gip}
\end{figure}

 \section{Coherent surface plasmon amplification}
 Applying the introduced model to the spaser implies that we assume the coupling between the quantum emitters and the plasmon mode to be identical for every quantum emitter.
 This is an approximation, but corresponds to  using a mean value for the coupling. 
 A coupling between the quantum emitters and plasmon particle can be derived e.g. using the dipole approach of Ref. \onlinecite{Ridolfo:PhysRevLett:10}. 
 This is a model assumption, but not unrealistic, since the quantum emitters are distributed randomly around the metal particle with a typical average distance to the metal nanoparticle \cite{Noginov:Nature:09,Bergman:PhysRevLett:03}. In addition, the metal particle plasmon resonance is spectrally  broad compared to the emission lines. Also  deviations from the mean quantum emitter frequency are of minor importance for the coupling to the plasmons. Different dipole orientations could play a role in the emitter plasmon interaction. Geometries where the surface plasmon and emitter dipole moments are parallel lead to the strongest couplings, therefore we believe the effects from these parallel contributions to be dominant. The magnitudes of all system parameters are discussed in Appendix \ref{app_param}.\\
 With the self-consistent theoretical framework, Eqs. (\ref{eq_mo_start}-\ref{eq_mo_end})  a thorough theoretical analysis of the plasmon-quantum emitter  system is carried out: we calculate the full time evolution of the system starting in the thermal equilibrium state at room temperature. The plotted quantities are steady state values of system observables as a function of the pump rate.
 We focus on the probabilty to find $k$ plasmons 
 \begin{equation}
 P_{pl}(k)=\sum_{n=0}^N {N \choose n} \rho_{[n,0,0,k,k]}
 \label{eq.plasmondist}
 \end{equation}
 or excitons 
 \begin{equation}
 P_{ex}(k)={N \choose k} \sum_{m} \rho_{[k,0,0,m,m]},
 \end{equation}
 in the system. Further quantites are the average number of plasmons 
 \begin{equation}
 \langle b^\dagger b\rangle=\sum_m m P_{pl}(m)
 \end{equation} 
 and excitons
 \begin{equation}
 n_C=\langle \mu_{ex}^\dagger \mu_{ex}\rangle/|\mu|^2=\sum_{n=1}^N n P_{ex}(n),
 \end{equation}
  as well as the plasmonic and exciton intensity-intensity correlation 
 \begin{equation}
 \langle b^\dagger b^\dagger b b\rangle=\sum_m m (m-1) P_{pl}(m)
 \end{equation}  
 and 
 \begin{equation}
 \langle \mu_{ex}^\dagger \mu_{ex}^\dagger \mu_{ex} \mu_{ex}\rangle/|\mu|^4=\sum_{n=1}^N n (n-1) P_{ex}(n)
 \end{equation}
 with $\mu_{ex}^\dagger=\mu\sum_i a^\dagger_{c_i} a_{v_i}$, where $\mu$ is the exciton dipole moment.
 
To understand these quantities as a function of the pump rate, one has to recognize that the dynamics of the plasmon quantum emitter system is governed by a strong  imbalance of dephasings of the material parameters of the different constituents of the spasers.
The plasmon dephasing is orders of magnitudes larger compared to the  emitter dephasing, regardless if the  emitter is a dye or quantum dot.
For the numerical evaluation, we assume a spherical metal nanoparticle and a spherical but random distribution of the quantum emitters in a surrounding shell, all parameters can be found in Appendix \ref{app_param}.
We choose parameters close to Ref. \onlinecite{Noginov:Nature:09}, but smaller metal nanoparticles and using silver instead of gold, a choice which actually should improve the possibility of spaser action, since it increases coupling and decreases dissipation. So all conclusions in the paper do apply for gold in a even more bonded way.
  
As a first step to analyze the operation of a spaser the average number of plasmons is calculated in dependence on the pump rate (Fig. \ref{inputoutput} a)).
We see that the average number of plasmons has a peak at intermediate pump rates and approaches zero for high pump rates, which is a known quenching effect of the incoherent pump term \cite{Gartner:PhysRevA:84}. The absence of a clear kink (transition from spontaneous to stimulated plasmon emission) in this input-output curve is a feature of a thresholdless  system \cite{Rice:PhysRevA:94} (note the logarithmic scale of Fig. \ref{inputoutput} a)). The thresholdless behaviour 
indicates  \cite{Rice:PhysRevA:94,Gartner:PhysRevA:84}
that  the 
input-output curve can not be used as indication for spasing, which is in agreement with the literature \cite{Protsenko:PhysUsp:12}.\\

This property should hold for all conceivable spaser parameters: from Eq. (\ref {liouville_vonneumann}) it is possible to derive rate equations for the spaser in complete analogy to the laser rate equations. Using equations similar to Ref. \onlinecite {Gies:PhysRevA:07} the Purcell enhancement of the spontaneous emission rate $\gamma =\gamma_x+\gamma_l$, which is included in our description, is set as 
\begin{equation}
\gamma_l=4 \frac{g^{2}}{\gamma_x+\gamma_{sp}+P+2\gamma_{pd}}.
\end{equation}
The $\beta$ factor is $\beta =(1+\gamma_x /\gamma_l)^{-1}$ showing that $\beta \sim 1$ holds, since for realistic spaser parameters $\gamma_l$ is magnitudes larger than $\gamma_x$.
   
As a second step, we discuss at what pump rates the emission in Fig. \ref{inputoutput} a) actually corresponds to coherent plasmons, i.e. spasing.
In Fig. \ref{inputoutput} b) the average number of excitons is plotted over the pump rate, which saturates for high pump rates at the number of quantum emitters.
We observe that the build up of the peak in the average plasmon number in  Fig. \ref{inputoutput} a) occurs in the regime in which the average number of excitons grows linearly (see Fig. \ref{inputoutput} b). In laser physics this is a well known indication for the onset of lasing\cite{Gartner:PhysRevA:84} -- consequently we can view it as an indication for spasing within the theory, which is however not observable in an experiment.

A sufficient condition for spasing is the presence of a Poissonian plasmon distribution.  
However, typically in experiments the determined quantity is the plasmon -- plasmon (photon -- photon) correlation function $g^{(2)}_{pl}$: it is defined as $g^{(2)}_{pl}(\tau=0)= \langle b^\dagger b^\dagger b b\rangle/\langle b^\dagger  b\rangle^2$, a value of $1$ suggests a coherent distribution - the Poissonian limit of $P_{pl}$, Eq. \eqref{eq.plasmondist} and a value of $2$ suggests a thermal distribution. 
It can be measured from photons emitted from the metal nanoparticle using a intensity-intensity correlation of photons  in a Hanbury-Brown Twiss experiment \cite{Brown:Nature:56}.
The $g^{(2)}_{pl}$ function is plotted as a function of the incoherent pump rate $P$ for $10$ and $30$ quantum emitters in Fig. \ref{g2gip} a) and b), respectively.
Before and after the onset of spasing the $g^{(2)}_{pl}$ has values near $2$ suggesting a thermal distribution (spontaneous emission of plasmons).
Whereas for intermediate pump rates, where linear increasing plasmon numbers in Fig. \ref{inputoutput} suggested spasing, $g^{(2)}_{pl}$ approaches $1$ indicating coherent plasmon emission. 
Comparing Figs. \ref{g2gip} a) and b) with Fig. \ref{inputoutput}, it is clear that the region of coherent plasmon states cannot be deduced from the input output curve for the spaser described for realistic experimental parameters.
Also the necessary pump rate for spasing is extremely high (2-3 orders of magnitudes higher than QD lasers \cite{Gies:PhysRevA:07}), due to the extremely high plasmon dephasing rate.
The necessary pump rates rather increase for higher numbers of quantum emitters instead of decreasing.
So  we conclude that it is probably very difficult to achieve  the spasing limit experimentally, which is in agreement with Refs. \onlinecite{Zhong:PhysRevB:13,Khurgin:NatPhoton:14}.
 
\begin{figure}
\includegraphics[clip,width=6cm]{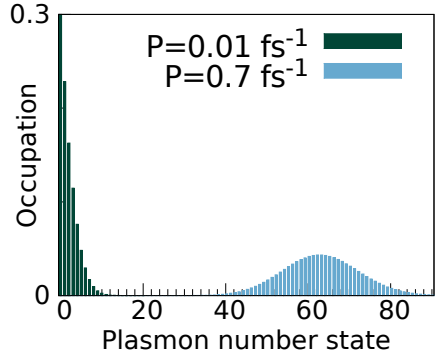}
\caption{A plot over the plasmon number distribution for two pump rates ($P=0.01$, $~0.7~\mathrm{fs}^{-1}$) before and after the onset of spasing.}
\label{plstatcomp}
\end{figure}

In general, a $g^{(2)}_{pl}$ of $1$ is only a necessary but not a sufficient criterion for a coherent state.
Therefore we discuss the full  plasmon distribution function before and after the spasing transition. In Fig. \ref{plstatcomp} b) we see that for pump rates above the spasing threshold the distribution $P_{pl}(n)$ (Eq. \eqref{eq.plasmondist}) changes from a thermal to a Poisson-like distribution as expected.  
  
Additionally to a $g^{(2)}_{pl}$ function of plasmon system, a $g^{(2)}_{ex}$ for the exciton system  $g^{(2)}_{ex}= \langle \mu_{ex}^\dagger \mu_{ex}^\dagger \mu_{ex} \mu_{ex}\rangle/\langle \mu_{ex}^\dagger  \mu_{ex}\rangle^2$  can be defined to the statistics  of the exciton system: the excitons behave like Bosons except that the corresponding Fock space is truncated at the number of emitters $N$, which can have severe influences on the statistical properties.
In Figs. \ref{g2gip} a) and b) $g^{(2)}_{ex}$ is plotted. The exciton $g^{(2)}_{ex}$ does not reach the coherent limit $g^{(2)}_{ex}=1$ in the spasing regime, it remains slightly above. The same is true before the spaser transition, suggesting deviations from classical light for light emitted from the excitons for a $g^{(2)}_{ex}>2$.
Anyway for an increasing number of emitters $g^{(2)}_{ex}$  is getting closer to $1$, suggesting coherent light emitted from the quantum emitters for very high emitter numbers. 
The behavior of $g^{(2)}_{ex}$ for increasing emitter numbers suggest Pauli blocking  as origin of the non bosonic properties of the emitter excitons  as source of the deviations from the coherent state.

\section{Summary and Conclusion}
In summary, we introduced a numerically exact method to handle a system of $N$ identical, externally driven quantum emitters coupled to a lossy optical mode. We applied the new formalism to the question of spasing (so far evaluated for $10$ and $30$ emitters) and found that although it is in principle possible it is very unlikely to be experimentally achieved and that the claims of realization in Ref. \onlinecite{Noginov:Nature:09} are probably incorrect. Our method is however not limited to the spaser, it constitutes a general and numerically exact solution to open many quantum emitter -- optical mode/cavity systems. In particular it does not rely on techniques such as adiabatic elimination or linearized fluctuations.

\begin{acknowledgments}
We gratefully acknowledge support from the Deutsche Forschungsgemeinschaft (DFG) through SPP 1391 (M.R.) and  SFB 951 (T.S.T., A.K.).
\end{acknowledgments}

\appendix
\section{Parameters and lifetimes}
\label{app_param}
Following the calculation \cite {Ridolfo:PhysRevLett:10} for
  a single emitter, we derive an averaged coupling assuming a spherical
  distribution $g=3/2\protect \sqrt {3\eta R^{3}/(\epsilon _0 \hbar )} \mu
  /(r_{2}^{3} - r_{1}^{3}) \protect \mathrm {ln}(r_{2}/r_{1}).$ Here $R$, $r_1$
  and $r_2$ are metal nanoparticle radius and inner and outer shell radii
  surrounding the nanoparticle, $\mu $ the quantum emitter transition dipole
  moment and $\eta =(\partial _{\omega } \epsilon '(\omega =\omega
  _{sp}))^{-1}$ the inverse derivative of the real part of the relative
  dielectric function $\epsilon '(\omega )$ at the dipole plasmon frequency
  $\omega _{sp}$. The Fr\IeC {\"o}hlich condition $\epsilon '(\omega
  _{sp})=-2\epsilon _h$ with host dielectric constant $\epsilon _h$ sets
  $\omega _{sp}$. The quantum emitter dipole moment is $\mu =0.7~\mbox{e~nm}$, gives a spontaneous emission rate of $\gamma
  _x=0.003~\mbox{ps}^{-1}$. The pure dephasing rate is
  $\gamma _{pd}=3~\mbox{ps}^{-1}$. We consider a small silver
  nanoparticle of $R=6~\mbox{nm}$, surrounded by a $6~\mbox{nm}$ shell, i.e. $r_1=6~\mbox{nm}$,
  $r_2=12~\mbox{nm}$, with dielectric constant of $\epsilon
  _h=3$, shifting the silver plasmon energy into the visible, resulting in a
  coupling strength and plasmon damping rate of $\hbar g=19.7~\mbox{meV}$ and $\gamma _{sp}=80~\mbox{ps}^{-1}$. The
  plasmon transition dipole moment is \protect \cite {Ridolfo:PhysRevLett:10}
  $\chi = \epsilon _h \protect \sqrt {12 \pi \epsilon _0 \hbar \eta R^{3}} =
  16.2 ~\mbox{e~nm}$.

\end{document}